\documentclass{elsarticle}

\usepackage{lineno,hyperref}
\usepackage{color}
\modulolinenumbers[5]

\journal{Journal of Chaos, Solitons \& Fractals}

\usepackage{amssymb}
\usepackage{amsmath}

\newcommand{\R}{\mathbb{R}}
\newcommand{\Q}{\mathbb{Q}}
\newcommand{\M}{\mathcal{M}}
\newcommand{\vol}{\operatorname{vol}}

\newtheorem{thm}{Theorem}

\newdefinition{rem}{Remark}
\newproof{pf}{Proof}
\begin{document}

\begin{frontmatter}

\title{Automatic Generation of Bounds for Polynomial Systems with Application to the Lorenz System\tnoteref{mytitlenote}}
\tnotetext[mytitlenote]{Partially supported by the German Academic Scholarship Foundation.}


\author[tud]{Klaus R{\"o}benack\corref{mycorrespondingauthor}}
\cortext[mycorrespondingauthor]{Corresponding author.}
\ead{klaus.roebenack@tu-dresden.de}

\author[tud,htwk]{Rick Vo{\ss}winkel}
\ead{rick.vosswinkel@htwk-leipzig.de}

\author[htwk]{Hendrik Richter}
\ead{hendrik.richter@htwk-leipzig.de}

\address[tud]{Technische Universit{\"a}t Dresden, Faculty of Electrical and Computer Engineering, Institute of Control Theory, 01062 Dresden, Germany}
\address[htwk]{HTWK Leipzig University of Applied Sciences, Faculty of Electrical Engineering and Information Technology,  04107 Leipzig, Germany}

\begin{abstract}
This study covers an analytical approach to calculate positive invariant sets of dynamical systems. Using Lyapunov techniques and quantifier elimination methods, an automatic procedure for determining bounds in the state space as an enclosure of attractors is proposed. The available software tools permit an algorithmizable process, which normally requires a good insight into the systems dynamics and experience. As a result we get an estimation of the attractor, whose conservatism only results from the initial choice of the Lyapunov candidate function.  The proposed approach is illustrated on the well-known Lorenz system.
\end{abstract}


\end{frontmatter}

\linenumbers

\section{Introduction}

A dynamical system may have an attractor which implies a neighborhood around the attractor on which all trajectories are bounded. Thus, it is interesting to ask if such a neighborhood can be described analytically.  From a geometric point of view, this means we intend to find a subset of the state space with some special properties, which is also called finding a compact invariant set or calculating an enclosure of the attractor. A standard procedure for calculating such a bound is to employ positive Lyapunov-like functions. However, there are two major restrictions to employing Lyapunov-like functions. A first is that the algebraic form as well as the parameterization of the function  offers a considerable degree of choice, which usually 
makes finding a suitable candidate a matter of (mostly human) trail-and-error. A second is that calculating symbolic or numerical values of the bound on the Lyapunov-like candidate function  that gives an estimate of the compact invariant set  commonly requires human insight, experience, and frequently substantial algebraic manipulations. 
The approach we present here aims at circumventing these problems by proposing an automatic and algorithmizable procedure.  This is done using quantifier elimination (QE) methods. The term quantifier elimination covers several methods \cite{collins1974quantifier,weispfenning1988complexity,caviness2012quantifier} to reformulate quantified formulas into a quantifier free equivalent. This idea has already been applied for stability analysis~\cite{hong1997stability,nguyen2003qe,she2009,vosswinkel2018ecc}, model verification~\cite{sturm2011verification} as well as controller design~\cite{sturm2011verification,jirstrand1997,dorato1998non}.

The method we propose calculates an analytic expression of compact positive invariant sets. Moreover, if a dynamical system is dissipative and may consequently possess one (or even several) attractors, then a positive invariant set may  contain at least one of them. However, attractors  are invariant sets with additional requirements. They have to be compact, they are not dividable into two invariant, disjoint subsets and they need an attractive neighborhood. This consequently applies to
chaotic attractors, which have a complicated geo\-metry, and for which compact positive invariant sets can provide an enclosure. Thus, calculating  attractor enclosures by compact positive invariant sets is related to, but differs in methodology and objective from computing analytic expressions of attractors themselves, as shown by calculating invariant measures and fractal dimension for the 2D Lorenz map~\cite{graca2017,galatolo2016}, or almost-invariant sets and invariant manifolds for the Lorenz system~\cite{froyland2009}. 
These differences in objective and methodology stem from this paper using  Lyapunov--like function  for calculating attractor enclosures, while calculating attractor approximations has been shown by either using a geometric description of the dynamics by invariant manifolds or a probabilistic description of dynamics by transfer operators or almost-invariants sets~\cite{araujo2014,graca2017,galatolo2014,galatolo2016,froyland2009}.

Although our method is generally applicable to dynamical systems with polynomial description, we specifically 
apply it to the Lorenz system to have a comparison with previous results. The Lorenz equations~\cite{lorenz63,sparrow1982} are arguable one of the best-known and most-studied dynamical systems that exhibit chaotic solutions. This also includes
several works on ultimate bounds, compacts sets, or attractor enclosures~\cite{sparrow1982,reitmann1987,krishchenko2006,li2005,leonov1987zamm-engl,suzuki2008,zhang16,pogromsky2003}.
 Apart from analyzing a property of the Lorenz system, calculating bounds is also a possible starting point for applications, for instance estimating the fractal dimension~\cite{boichenko2005dimension} or the Hausdorff dimension of the Lorenz attractor~\cite{pogromsky2003}. Attractor enclosures have additionally been used for the tracking of periodic solutions, stabilization of equilibrium points and synchronization~\cite{pogromsky2003,richter2001,yu2006globally}.

The paper is structured as follows: In Section~\ref{sec:dyn-eq} we introduce our approach with briefly recalling quantifier elimination and calculating bounds on trajectories using Lyapunov-like functions. We also discuss how quantifier elimination can be used to obtain such bounds.
The method is applied to the Lorenz system in Section~\ref{sec:lorenz}. We calculate spherical and elliptical bounds with fixed and variable center points and show that  our method  can be used to reproduce, algebraically verify and partly improve bounds known from previous works~\cite{sparrow1982,li2005,leonov1987zamm-engl}. In Section~\ref{sec:cons} we derive some conclusions.

\section{Computation of Bounds for Dynamical Systems by Quantifier Elimination}
\label{sec:dyn-eq}

\subsection{Real Quantifier Elimination}
\label{sub:QE}

Before we illustrate the proposed method let us briefly introduce some mathematical preliminaries on quantifier elimination (QE), cf.~\cite{collins1975quantifier,caviness1998}, starting with a simple example to delineate the main ideas of QE.

Let us consider the quadratic function $g(x)=a_2x^2+a_1x+a_0$. The question if a parameter constellation $u=(a_0,a_1,a_2)$ exists such that the function values $g(x)$ are always positive can be formulated using the quantified expression
\begin{equation*}
\exists\, a_2,a_1,a_0 ~\forall x: g(x)>0,
\end{equation*}
which can easily be answered with \texttt{true}. If we are interested in all parameter constellations~$u$, which result in $g(x)>0$, we utilize QE. Therefore, we omit the quantifiers for~$u$ to generate an equivalent expression in these quantifier-free variables
\begin{equation*}
\forall x: g(x)>0.
\end{equation*}
Applying a QE method to the problem we get 
\begin{equation*}
(a_1=0 \lor 4a_2a_0-a_1^2 \neq 0)\land a_0>0 \land -4a_2a_0+a_1^2\leq 0.
\end{equation*}
Thus,  we get exact conditions which are equivalent to the previous formula.
After presenting the necessary fundamentals of QE, it is next shown how these techniques can be applied to estimate positive invariant sets.

In the following, we introduce the concept of quantifier elimination in a more formal way.
An \emph{atomic formula} is an expression of the form
\begin{equation}
\label{eq:atomic}
 \phi(x_1,\ldots,x_k) \;\tau \;0
\end{equation}
with a relation $\tau \in\{>,=\}$, where $\phi\in\Q[x_1,\ldots,x_k]$ is a polynomial in the variables $x_1,\ldots,x_k$ with rational coefficients. A combination of atomic formulas~\eqref{eq:atomic} with the Boolean operators $\land, \lor, \lnot$ is called a \emph{quantifier-free formula}. With these standard operators we can express all other Boolean operators such as equivalence ($\iff$) or implication ($\implies$) and augment the list of relations for~\eqref{eq:atomic} to $\{<,\leq,>,\geq,=,\neq\}$.

Let $F(u,v)$ be a quantifier-free formula in the variables $u=(u_1,\ldots,u_k)$ and $v=(v_1,\ldots,v_l)$. 
A \emph{prenex formula} is an expression 
\begin{equation}
 \label{eq:prenex}
 G(u,v):=(Q_1 v_1)\cdots (Q_l v_l)\,F(u,v)
\end{equation}
with quantifiers $Q_i\in\{\exists,\forall\}$ for $i=1,\ldots,l$. The variables~$v$ are called \emph{quantified} and the variables~$u$ are called \emph{free}, respectively. Thus, the parameters $\{ a_2,a_1,a_0\}$ gives the set $u$ and $\{x\}$ gives the set $v$ in the before described example of the quadratic equation.
The quantifiers occurring in~\eqref{eq:prenex} can be eliminated~\cite{tarski1948decision,tarski1998decision,Seidenberg1954}. This process is referred to as \emph{quantifier elimination}. The following theorem is a direct consequence of the well-known Tarski-Seidenberg-Theorem~\cite[pp. 69-70]{basu2006}:

\begin{thm}[Quantifier Elimination over the Real Closed Field]
\label{thm:tarski}
For every prenex formula $G(u,v)$ there exists an equivalent quantifier-free formula $H(u)$.
\end{thm}

The first algorithm to determine such a quantifier-free equivalent was presented by Tarski itself. Unfortunately, this algorithm was not applicable because its computational complexity can not be bounded by any stack of exponentials. The first procedure which could be applied to non-trivial problems is \emph{cylindrical algebraic decomposition} (CAD)~\cite{collins1975quantifier}. This algorithm mainly consists of four steps. The first decompose the space in so-called cells in which every polynomial has a constant sign. Secondly, these cells are gradually projected from $\R^n$ to $\R^1$. These projections are cylindrical and algebraic. The conditions of interest are evaluated in $\R^1$ in the third step and the results are finally lifted to $\R^n$. Due to the universal applicability to the input sets of polynomials, this algorithm and its improvements (e.g. \cite{collins1991}) are still often used. Nevertheless, in the worst case  the computational effort is doubly exponential in the number of variables~\cite{davenport1988}. 

The second commonly used procedure is \emph{virtual substitution}~\cite{weispfenning1988complexity,Loos1993,weispfenning1994}. The problem $\exists\,v: F(u,v)$ is solved with a formula substitution equivalent, where $a$ is substituted with terms of an elimination set. This procedure is just applicable to linear, quadratic and cubic polynomials, but the resulting complexity is "just" exponential in the number of quantified variables. Furthermore,  the resulting conditions are often very large and redundant such that a subsequent simplification is necessary. 

A third frequently applied method for QE is based on \emph{real root classification} (RRC). The number of real roots in a given interval can be computed using Sturm or Sturm-Habicht sequences. Based on that idea, formulations to eliminated quantifiers can be generated \cite{Gonzalez1989, Yang1996, Iwane2013}. As in the case of virtual substitution the resulting output formulas are often very large and redundant such that a subsequent simplification is need as well. However, very effective algorithms can be achieved, especially for sign definite conditions $\forall v\geq 0 \implies f(u,v)>0$, see~\cite{Iwane2013}. 

To carry out the quantifier elimination we used the open-source software packages QEPCAD~\cite{collins1991,brown2003qepcad},
and REDLOG~\cite{dolzmann1997redlog}. The later package is part of the computer algebra system REDUCE.
For both tools, the resulting quantifier-free formulas can be simplified with the tool SLFQ~\cite{brown2006slfq}.
The computations were carried out on a standard PC with Intel\textsuperscript{\textregistered} Core\texttrademark\ i3-4130 CPU at 3.4\,GHz and 32\,GiB RAM under the Linux system Fedora~25 (64\,bit).
For QE we used the advanced virtual substitution method from~\cite{kosta2016} (i.e., function \texttt{rlqe} with the switch \texttt{on ofsfvs}).
The authors made the source code of prototype implementations publically available on Github~\cite{github-bounds-lorenz} under the GNU GPL v3.0
in order to allow a verification of the presented results.

\subsection{Bounds in Terms of a Lyapunov-like Function}

Consider an autonomous nonlinear system
\begin{equation}
 \label{eq:sys}
 \dot{x}=f(x)
\end{equation}
with the vector field $f:\M\to\R^n$ defined on an open subset $\M\subseteq\R^n$.
We assume that~\eqref{eq:sys} has a global solution $x(\cdot)$ for all $x(0)\in\M$.
Bounds on system~\eqref{eq:sys} are often formulated in terms of a continuously differentiable Lyapunov-like function $V:\M\to[0,\infty)$ and a constant $\gamma>0$ as
\begin{equation}
 \label{eq:bound-def}
 \limsup_{t\to\infty} V(x(t))\leq\gamma.
\end{equation}
The computation of the bound~$\gamma$ by~\eqref{eq:bound-def} would require the knowledge of the solution~$x(\cdot)$, which is generally not available for nonlinear systems. However, a bound~$\gamma$ could be obtained using Lyapunov techniques by
\begin{equation}
 \label{eq:bound-impl}
 \forall x\in\M:\quad V(x)>\gamma\;\implies\;\dot{V}(x)<0.
\end{equation}
This means that the subset $\mathcal{E}:=\{x\in\M:\;V(x)\leq\gamma\}$ is positive invariant, i.e., for all initial values $x(0)\in\mathcal{E}$ we have $x(t)\in\mathcal{E}$ for all $t\geq 0$.
Similarly, the bound~$\gamma$ could be obtained from
\begin{equation}
 \label{eq:bound-ode}
 \exists \alpha>0\;
 \forall x\in\M :\quad \dot{V}(x)\leq -\alpha\cdot (V(x)-\gamma),
\end{equation}
where trajectories starting from $x(0)\notin\mathcal{E}$ converge to~$\mathcal{E}$.

In general, the computation of~$\gamma$ from~\eqref{eq:bound-impl} or~\eqref{eq:bound-ode} seems to be difficult since these formulas contain quantifiers.  If~$V$ and~$\dot{V}$ are polynomials we could employ quantifier elimination  in order to obtain a quantifier-free expression on the bound~$\gamma$. As mentioned in Section~\ref{sub:QE},  quantifier elimination is an algorithmizable process that removes the universal quantifier $\forall$ and the existential quantifier  $\exists$ from the expressions \eqref{eq:bound-impl} or \eqref{eq:bound-ode}. This process can be seen as a simplification as a quantifier-free formula for \eqref{eq:bound-impl} or \eqref{eq:bound-ode} may be obtained that can subsequently be evaluated automatically to get the bound $\gamma$. Within the restriction of computational feasibility, the procedure can even be extended to not only including the bound $\gamma$ but selected parameters of the candidate function $V$, for instance the center points of spherical invariant sets or the center points and axes of elliptical ones. Thus, the quantifier-free formula can be evaluated w.r.t. bounds \emph{and} parameters of the Lyapunov-like function $V$.  \\ 

\section{Application to the Lorenz System}
\label{sec:lorenz}

We apply our approach to the Lorenz system~\cite{lorenz63}
\begin{equation}
\label{eq:lorenz}
\begin{array}{lcl}
\dot{x}_1&=&s(x_2-x_1)\\
\dot{x}_2&=&r\,x_1-x_2-x_1x_3\\
\dot{x}_3&=&x_1x_2-bx_3
\end{array}
\end{equation}
with positive parameters $s,r,b>0$. In some cases we will use the parameter values
\begin{eqnarray}
\label{eq:lorenz-param}
s=10,\quad
r=28,\quad
b=8/3.
\end{eqnarray}

\subsection{Spherical Bounds with Fixed Center}
\label{sub:sphere-fixed}

\begin{figure}
 \begin{center}
  \includegraphics{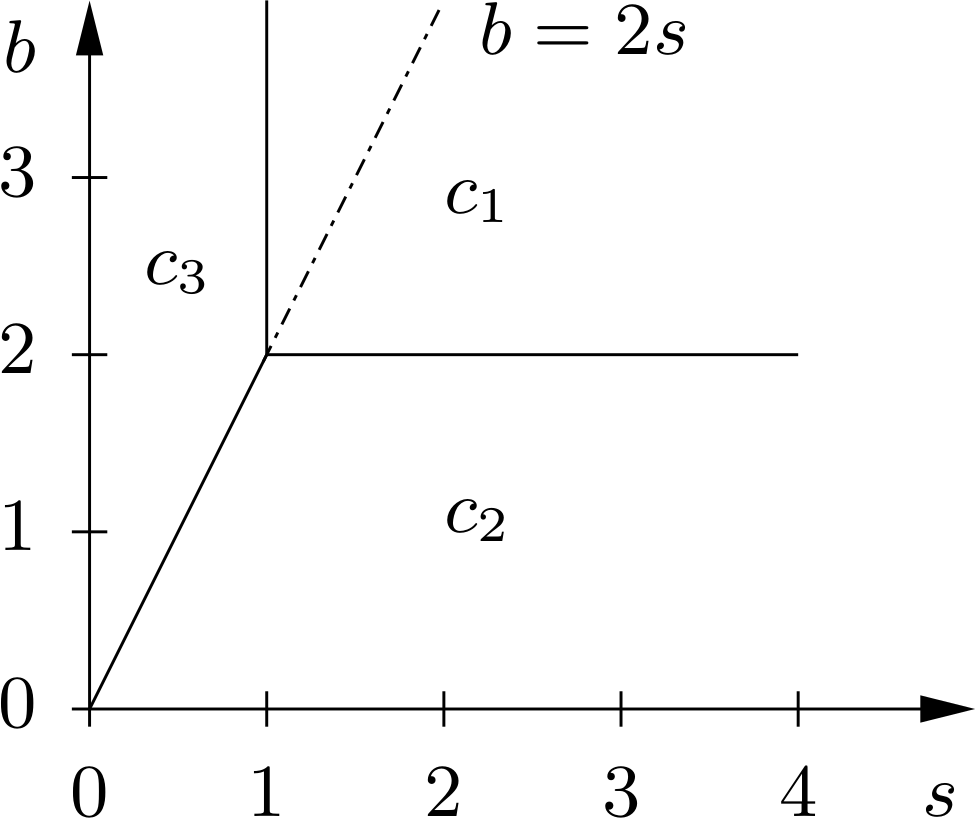}
 \end{center}
\caption{Parameter space for the bounds given in Th.~\ref{thm:li2005}}
\label{fig:parameter-space}
\end{figure}

First, we verify our approach using a well-known bound given in~\cite{li2005} based on the
Lyapunov-like function
\begin{equation}
 \label{eq:V1}
 V_1(x)=x_1^2+x_2^2+(x_3-r-s)^2.
\end{equation}
We consider level sets of~\eqref{eq:V1} with $\gamma=c^2$ with $c>0$.
Geometrically, these level sets are spheres with the radius~$c$ around the fixed center point $(0,0,r+s)$.
Defining
\begin{equation}
\label{eq:c123}
 c_1:=\frac{(s+r)b}{2\sqrt{b-1}}, \;
 c_2:=r+s,\;
 c_3:=\frac{(s+r)b}{2\sqrt{s(b-s)}}
\end{equation}
the results presented in~\cite[Th.~2]{li2005} can be stated as follows:

\begin{thm}
\label{thm:li2005}
A bound of \eqref{eq:lorenz} for $V_1(x)\leq c^2$ is given by
\begin{equation}
\label{eq:c-li2005}
 c=\left\{
 \begin{array}{lll}
 c_1& \text{for}& s\geq1 \land b\geq2, \\
 c_2& \text{for}& 2s>b \land b< 2,\\
 c_3& \text{for}& 2s\leq b \land s<1.
 \end{array}
\right.
\end{equation}
\end{thm}
The associated partition of the parameter space $(s,b)$ is sketched in Fig.~\ref{fig:parameter-space}.
The first case was already derived in~\cite{leonov1987zamm-engl} with $c=c_1=152/\sqrt{15}\approx39.246$ for~\eqref{eq:lorenz-param}.

Using the Lyapunov-like function~\eqref{eq:V1} we will derive bounds based on quantifier elimination.
In our case, Eq.~\eqref{eq:bound-impl} has the form
\begin{equation}
 \label{eq:V1-quant}
 \forall x_1,x_2,x_3:\quad s,r,b,c>0 \land [V(x)>c^2\,\implies\,\dot{V}(x)<0 ]
\end{equation}
with the quantified variables $x_1,x_2,x_3$. To obtain a quantifier-free representation of~\eqref{eq:V1-quant} w.r.t.\ the free variables $s,r,b,c$ we used the package REDLOG~\cite{dolzmann1997redlog} with the computer algebra system REDUCE. We obtained expressions taking 4,3\,KiB as plain ASCII code. 
Overall, the computation required approximately $250$\,ms computation time on the above mentioned platform, where about $100$\,ms where needed for quantifier elimination and about $150$\,ms to execute the REDLOG script and to store the result.
Simplifying these expressions with SLFQ~\cite{brown2006slfq} results after 130 QEBCAD calls in
\begin{subequations}
\label{eq:c-quant1}
\begin{align}
c\geq r+1 \;\land\; [&(b<2s \;\land\; b<2) \;\lor\\
 &(\text{Ineq.~\eqref{eq:in1}} \;\land\; b<2s) \;\lor\\
 &(\text{Ineq.~\eqref{eq:in2}} \;\land\; b<2) \;\lor\\
 &(\text{Ineq.~\eqref{eq:in1}} \;\land\; \text{Ineq.~\eqref{eq:in2}}) ]
\end{align}
\end{subequations}
with the inequalities
\begin{eqnarray}
\label{eq:in1}
4(b-1)c^2 &\geq& b^2(r+s)^2,\\
\label{eq:in2}
4s(b-s)c^2 &\geq& b^2(r+s)^2.
\end{eqnarray}
Clearly, the Ineqs.~\eqref{eq:in1} and~\eqref{eq:in2} correspond to the bounds~$c_1$ and~$c_3$ given in~\eqref{eq:c123}. Taking the different regions of the parameter space $(s,b)$ into account yields the bounds~\eqref{eq:c-li2005} given in Th.~\ref{thm:li2005}. 

\begin{rem}
With our formal approach, we obtained the same bounds as in ~\cite[Th.~2]{li2005}  with just a slightly different formulation.
Since~\eqref{eq:c-quant1} is  equivalent to the quantified expression~\eqref{eq:V1-quant}, the bounds given in Th.~\ref{thm:li2005} are strict w.r.t.\ the levels sets of the Lyapunov function~\eqref{eq:V1}, i.e., with~\eqref{eq:V1} the bounds~\eqref{eq:c-li2005} cannot be improved.
\end{rem}

\subsection{Spherical Bounds with Variable Center}
\label{sub:sphere-variable}

\begin{figure}
 \begin{center}
  \includegraphics[width=0.6\textwidth]{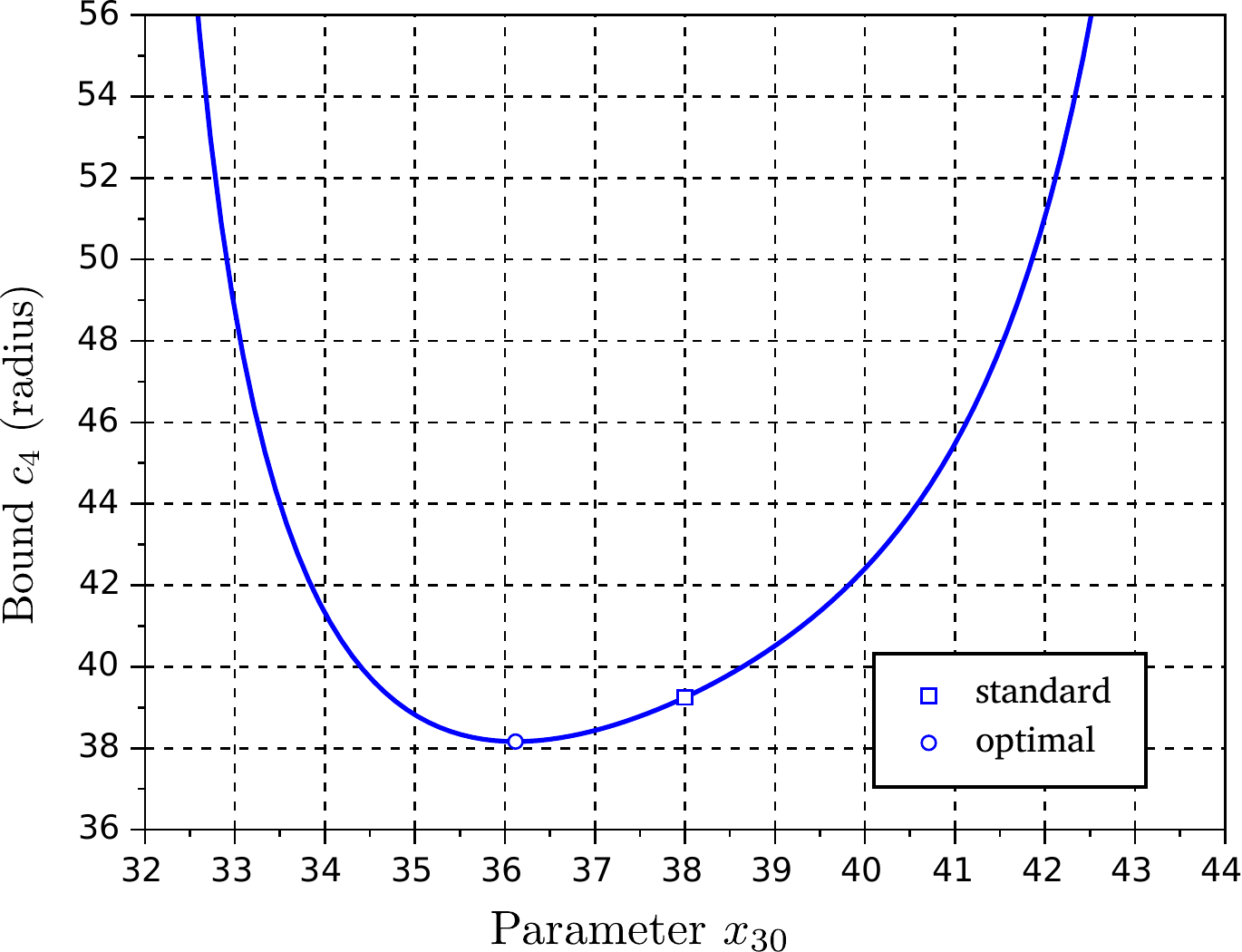} 
 \end{center}
\caption{Bound~$c_4$ depending on the parameter~$x_{30}$}
\label{fig:c4-radius}
\end{figure}

Now, we consider levels sets of spheres around $(0,0,x_{30})$ with a variable displacement~$x_{30}$ along the $x_3$-axis. This results in the Lyapunov-like function
\begin{equation}
 \label{eq:V2}
 V_2(x)=x_1^2+x_2^2+(x_3-x_{30})^2.
\end{equation}
The quantifiers in the associated expression~\eqref{eq:bound-impl} could not be eliminated using REDUCE. 
Therefore, we turned our attention to~\eqref{eq:bound-ode} yielding
\begin{equation}
 \label{eq:V2-quant}
 \exists \alpha \;\forall x_1,x_2,x_3:\quad \alpha,s,r,b,c>0 \land \dot{V}(x)\leq -\alpha\cdot (V(x)-c^2)
\end{equation}
with the free parameters $\alpha,s,r,b,c, x_{30}$. The quantifier elimination with REDUCE using variable (e.g.\ unspecified) parameters results in 868\,KiB
ASCII text for the equivalent quantifier-free expression requiring a computation time of approximately $11.4$\,s (including output and storage of the result).
Unfortunately, we were not able to simplify these expressions with SLFQ. However, with fixed parameters~\eqref{eq:lorenz-param} we obtained 
37\,KiB ASCII code in about $1.1$\,s CPU time, which could be simplified with SFLQ 
after about 500 calls of QEPCAD (depending on the options)
to
\begin{equation}
\label{eq:c-quant2}
x_{30}^2-76x_{30}+1404<0\;\land\;
 (c\geq{c}_4\;\text{ with }\;a_2{c}_4^4+a_1 {c}_4^2+a_0=0)
\end{equation}
with the coefficients
\[\begin{array}{lcl}
 a_0&=&4096 x_{30}^{4},\\
 a_1&=&384x_{30}^2(3x_{30}^2-228x_{30}+4762),\\
 a_2&=&9(x_{30}^2-76x_{30}+1404)(9x_{30}^2-684x_{30}+13436).
\end{array}
\]
The first inequality in~\eqref{eq:c-quant2} yields the open interval
\begin{equation}
\label{eq:intervall-x30}
 31.676 \approx 38-2\sqrt{10}<x_{30}<38+2\sqrt{10} \approx 44.524.
\end{equation}
From the second term in~\eqref{eq:c-quant2} we can compute the bound~$c_4$ on~$c$ for a given point~$x_{30}$ using the largest real root. The result is shown in Fig.~\ref{fig:c4-radius}. We have poles at the boundaries of the interval~\eqref{eq:intervall-x30} since $a_2\to0$ for $x_{30}\to38\pm2\sqrt{10}$.

Clearly, the choice $x_{30}=r+s=38$ from Section~\ref{sub:sphere-fixed} with~\eqref{eq:lorenz-param} is the center of the interval~\eqref{eq:intervall-x30}. However, the minimum of the bound~$c$ occurs at a slightly different point as shown in Fig.~\ref{fig:c4-radius}. The solution of the biquadratic equation in~\eqref{eq:c-quant2} can be computed symbolically. The critical points are determined using the first derivative test.
From~\eqref{eq:c-quant2} we compute the minimum $x_{30}\approx36.118$ with $c_{4}\approx38.1636$, which  is an improvement compared to the radius computed in Section~\ref{sub:sphere-fixed}.  

\subsection{Elliptical Bounds with Fixed Axes and Fixed Center}
\label{sub:ellipse-fixed}

Another well-known bound is derived in~\cite[Appendix~C]{sparrow1982} using the Lyapunov-like function
\begin{equation}
 \label{eq:V3}
 V_3(x)=rx_1^2+sx_2^2+s(x_3-2r)^2.
\end{equation}
The levels sets of~\eqref{eq:V3} are ellipsoids around the center point $(0,0,2r)$ with a specific scaling of the axes.
Quantifier elimination applying to the associated formula~\eqref{eq:bound-impl} directly yields
\begin{subequations}
\label{eq:c-quant3}
\begin{align}
c^2\geq 4sr^2 \;\land\; [&(b<2s \;\land\; b<2) \;\lor\\
 &(\text{Ineq.~\eqref{eq:in3}} \;\land\; b<2s) \;\lor\\
 &(\text{Ineq.~\eqref{eq:in4}} \;\land\; b<2) \;\lor\\
 &(\text{Ineq.~\eqref{eq:in3}} \;\land\; \text{Ineq.~\eqref{eq:in4}}) ],
\end{align}
\end{subequations}
with the inequalities
\begin{eqnarray}
\label{eq:in3}
(b-1)c^2 &\geq& sb^2r^2,\\
\label{eq:in4}
(b-s)c^2 &\geq& b^2r^2.
\end{eqnarray}
Taking the different regions of the parameters $b,s$ into account this leads immediately to the following theorem, which is similar to~\cite[Appendix~C]{sparrow1982}:
\begin{thm}
\label{thm:sparrow}
A bound of \eqref{eq:lorenz} for $V_3(x)\leq c^2$ is given by
\begin{equation}
\label{eq:c-sparrow}
 c=\left\{
 \begin{array}{lll}
 br\sqrt{\frac{s}{b-1}}& \text{for}& s\geq1 \land b\geq2, \\
 2r\sqrt{s}& \text{for}& 2s>b \land b< 2,\\
 \frac{br}{\sqrt{b-s}}& \text{for}& 2s\leq b \land s<1.
 \end{array}
\right.
\end{equation}
\end{thm}
For the parameters~\eqref{eq:lorenz-param} we obtain $c\approx 182.895$.

\subsection{Elliptical Bounds with Fixed Axes and Variable Center}
\label{sub:ellipse-variable}

\begin{figure}
  \includegraphics[width=\textwidth]{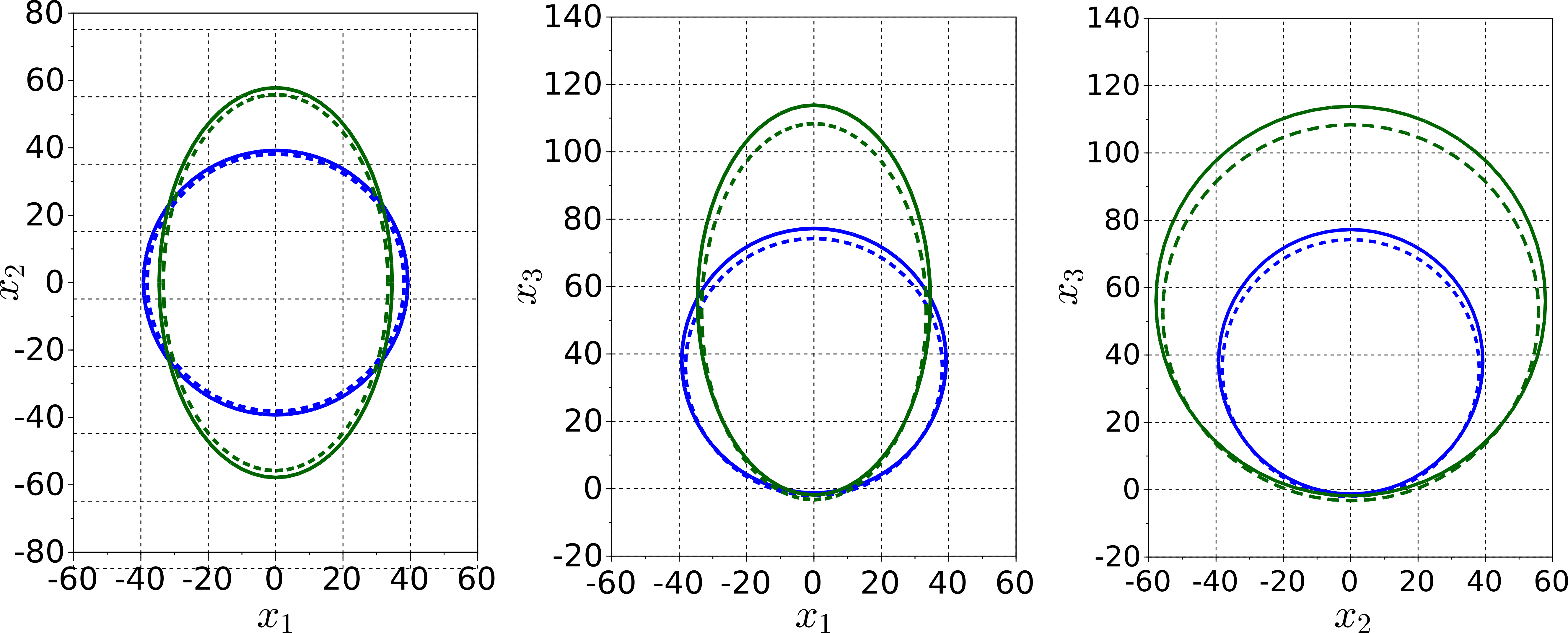}
\caption{Projections of the bounds 
from Eq.~\eqref{eq:V1} (blue, solid line),
Eq.~\eqref{eq:V2} (blue, dashed line),
Eq.~\eqref{eq:V3} (green, solid line),
Eq.~\eqref{eq:V4} (green, dashed line), respectively,
into two-dimensional planes}
\label{fig:bounds}
\end{figure}

Similar as in Section~\ref{sub:sphere-variable} we replace the fixed center point of~\eqref{eq:V3} by a variable one $(0,0,x_{30})$ and use the parameter set~\eqref{eq:lorenz-param}. This leads to the Lyapunov-like function
\begin{equation}
 \label{eq:V4}
 V_4(x)=rx_1^2+sx_2^2+s(x_3-x_{30})^2.
\end{equation}
From the quantifier elimination with REDUCE we obtained 40\,KiB ASCII source code in about $1.1$\,s computation time, that could be reduced with SLFQ after about 500 QEPCAD calls to the quantifier-free formula
\begin{equation}
\label{eq:c-quant4}
x_{30}>0\;\land\;
 (c\geq{c}_5\;\text{ with }\;a_2{c}_5^4+a_1 {c}_5^2+a_0=0),
\end{equation}
with the coefficients
\[
\begin{array}{lcl}
 a_0&=&3211264 x_{30}^4,\\
 a_1&=&10752 x_{30}^2 (3 x_{30}^2-336 x_{30}+10612),\\
 a_2&=&9 (x_{30}^2-112 x_{30}+3024) (9 x_{30}^2-1008 x_{30}+29456).
\end{array}
\]
The biquadratic equation in~\eqref{eq:c-quant4} has only for
\begin{equation}
\label{eq:intervall-ellipse}
 45.417\approx 56-4\sqrt{7}<x_{30}<56+4\sqrt{7} \approx 66.583
\end{equation}
 real roots. As in Section~\ref{sub:sphere-variable}, the center of the interval is $x_{30}=2r=56$ as used in Section~\ref{sub:ellipse-fixed}. Again, the minimum of~$c$ occurs not at the center of~\eqref{eq:intervall-ellipse} but at $x_{30}\approx 52.553$ with $c\approx176.531$.
The four bounds computed in Sections~\ref{sub:sphere-fixed} to~\ref{sub:ellipse-variable} are depicted in Fig.~\ref{fig:bounds} as projections into the different two-dimensional planes.

\subsection{Elliptical Bounds with Variable Axes and Center}
\label{sub:ellipse-general}

The quadratic Lyapunov-like functions $V_1,\ldots,V_4$ vary w.r.t.\ the scaling of the axes and the center point in the coordinate~$x_3$. Generalizing these functions results in
\begin{equation}
 \label{eq:V5}
 V_5(x)=p_1x_1^2+p_2x_2^2+p_3(x_3-x_{30})^2,
\end{equation}
with $p_1,p_2,p_3>0$ and $x_{30}\in\R$. 
The large expression sizes discussed in Sections~\ref{sub:sphere-variable} and~\ref{sub:ellipse-variable} suggest that we will not be able to solve the associated prenex formulas~\eqref{eq:bound-impl} or~\eqref{eq:bound-ode} with variable parameters. In addition to the system parameters~\eqref{eq:lorenz-param} we used pre-determined parameters $p_1,p_2,p_3,x_{30},c$ resulting in a prenex formula without free variables. Then, the equivalent quantifier-free formula (obtained after quantifier elimination) is either \texttt{true} or \texttt{false}. 

For given parameters $p_1,p_2,p_3,x_{30}$ we implemented a bisection method to approximately determine the boundary of~$c$. We carried out a Monte Carlo simulation with uniformly distributed parameters $p_1,p_2,p_3\in[0.1,5]$ and $x_{30}\in[10,80]$.  We did not find any admissible solution with $p_2\neq p_3$. 
This observation can be illustrated by the  derivative 
\begin{equation}
\label{eq:V5dot}
\begin{array}{rcl}
\dot{V}_5&=&\;2(p_3-p_2)x_1x_2x_3+(56p_2+20p_1-2p_3x_{30})x_1x_2 \\ 
&&-\frac{16}{3}p_3x_3(x_3-x_{30})-20p_1x_1^2-2p_2x_2^2. 
\end{array}
\end{equation}
The first term in~\eqref{eq:V5dot} needs to be compensated to achieve negative definiteness for $\|x\|\gg0$. 
This consideration leads to $p_2=p_3$ and restricts the set a suitable ellipsoids. 

The ellipsoid given by $V_5(x)\leq c^2$ has the volume
\begin{equation}
 \vol=\frac{4\pi}{3} \cdot \frac{c^3}{\sqrt{p_1 p_2 p_3}}.
\end{equation}
Tab.~\ref{tab:volumes} shows the bounds~$c$ and the volumes of the associated ellipsoids for different parameters $p_1,p_2,p_3,x_{30}$, where the Lyapunov-like functions $V_1,\ldots,V_4$ are considered as special cases of~$V_5$.
For~\eqref{eq:V5} we used $p_1=1$ and $p_2=p_3$. In order to minimize the volume of the positive invariant ellipsoid we carried out a Monte Carlo simulation. With this approach we were able to reduce the volume 
compared to the results given in Sections~\ref{sub:sphere-fixed} to~\ref{sub:ellipse-variable}.
\begin{table}
 \caption{Volumes of the ellipsoids for different parameter sets of function~\eqref{eq:V5}}
 \label{tab:volumes}
 \begin{center}
 \renewcommand{\arraystretch}{1.1}
 \begin{tabular}{cccccc}
  \hline 
  $V$ & $p_1$ & $p_2=p_3$ & $x_{30}$ & $c$ & $\vol$\\
  \hline
  $V_1$ & $1$ & $1$ & $38$ & $39.246$ & $2.532\cdot10^5$\\
  $V_2$ & $1$ & $1$ & $36.1177$ & $38.164$ & $2.328\cdot10^5$\\
  $V_3$ & $28$ & $10$ & $56$ & $182.895$ & $4.843\cdot10^5$\\
  $V_4$ & $28$ & $10$ & $52.563$ & $176.531$ & $4.355\cdot10^5$\\
  $V_5$ & $1$ & $1.62$ & $32.83$ & $43.956$ & $2.196\cdot10^5$\\
  \hline
 \end{tabular}
 \end{center}
\end{table}

\begin{figure}
  \includegraphics[width=\textwidth]{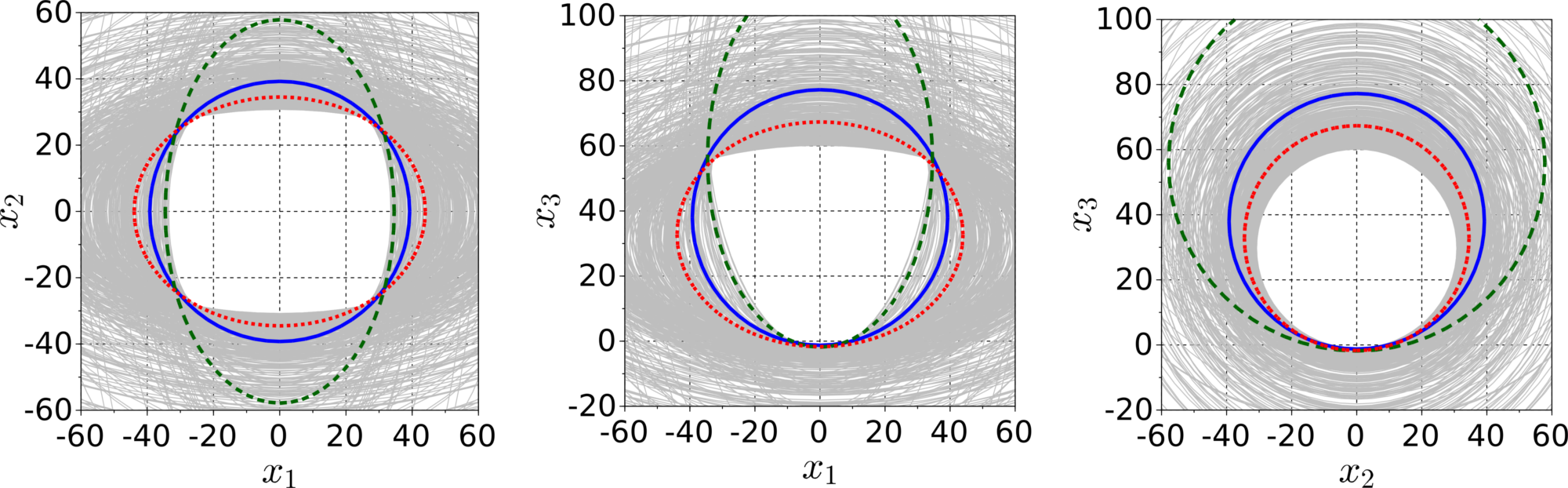} 
\caption{Bounds resulting from the union of 500 ellipsoids described by~\eqref{eq:V5} (gray, thin lines), 
bound resulting from~\eqref{eq:V1} with $c\approx39.246$ (blue, bold solid line) and bound resulting from~\eqref{eq:V3} with $c\approx182.895$ (green, long dashes),
best bound resulting from~\eqref{eq:V5} by Monte Carlo simulation (red, short dashes)}
\label{fig:union}
\end{figure}

After calculating several parameter constellations the union of the resulting ellipsoids gives a better approximation of the bounds then each individual one. This is shown in Figure~\ref{fig:union}. The union set can formally be described by the logical conjunction of the inequalities describing the ellipsoids.

\section{Conclusions}
\label{sec:cons}

The proposed approach gives a formal procedure to approximate the compact invariant set of nonlinear dynamical systems based on quantifier elimination methods. The resulting conservatism did not arise out of the calculation process itself but rather by the initial choice of the Lyapunov candidate. It was exemplarily shown how QE tools can be used to regenerate  and improve known results in an automatic manner. The most challenging aspect are the inherent computational barriers of QE methods. To reduce the computational effort to a manageable scale, it is often necessary to preprocess or decompose the problem. Nevertheless,  the proposed approach and quantifier elimination in general are very powerful and universal tools in system analysis and control. 

In this paper we used quadratic Lyapunov-like functions with ellipsoids as level sets. With quadratic functions we could also describe cylinders or elliptic cones. Our approach is not restricted to quadratic Lyapunov-like functions, i.e., we could use high order polynomials. 
The Lorenz system discussed in the paper is a good example of a system with complicated dynamics having comparatively simple bounds. From our experience, complicated dynamics itself (such the existence of a chaotic attractor) may not necessarily be an obstacle for our approach.
More important for a successful calculation is the relation between the systems dynamics and the geometry of the bound. In addition, higher order terms in the Lyapunov-like function can tighten the bound but may increase the computational effort significantly.
Our method can be used to derive bounds for other nonlinear systems, e.g. for the Lorenz-Haken system.

%
%
%



\section*{References}


\end{document}